\documentclass[aps,prc,groupedaddress,preprint,showpacs,amsmath,amssymb]{revtex4}
\usepackage{graphicx}
\usepackage{dcolumn}
\usepackage{bm}

\begin{document}

\title{Effects of pion potential and nuclear symmetry energy on the $\pi^{-}/\pi^{+}$ ratio in
    heavy-ion collisions at beam energies around the pion production
    threshold}

\author{Wen-Mei Guo$^{1,2,3}$}
\author{Gao-Chan Yong$^{1,4,5}$}\email{yonggaochan@impcas.ac.cn}
\author{Hang Liu$^{6}$}
\author{Wei Zuo$^{1,4,5}$}

\affiliation{%
$^1${Institute of Modern Physics, Chinese Academy of Sciences, Lanzhou 730000, China}\\
$^2${School of Physical Science and Technology, Lanzhou University,
Lanzhou 730000, China}\\
$^3${University of Chinese Academy of Sciences, Beijing 100049, China}\\
$^4${State Key Laboratory of Theoretical Physics, Institute of
Theoretical Physics, Chinese Academy of Sciences, Beijing, 100190}\\
$^5${Kavli Institute for Theoretical Physics, Chinese Academy of
Sciences, Beijing 100190, China}\\
$^6${Texas Advanced Computing Center (TACC) University of Texas at
Austin, Austin, Texas 78758, USA}\\
}%

\date{\today}

\begin{abstract}
Within the framework of the isospin-dependent
Boltzmann-Uehling-Uhlenbeck(IBUU) transport model, we studied the
effects of the pion potential and the symmetry energy on the pion
production in the central $^{197}Au+^{197}Au$ collisions around
the pion production threshold. It is found that the pion potential
has opposite effects on the value of $\pi^-/\pi^+$ ratio at low
and high pion energies. The effect of the pion potential on the
total $\pi^-/\pi^+$ ratio becomes larger in heavy-ion collisions
at beam energies below the pion production threshold. And at beam
energies below the pion production threshold, with the pion
potential, the effect of the symmetry energy on the $\pi^-/\pi^+$
ratio becomes smaller compared with that above the pion production
threshold.

\end{abstract}


\maketitle

\section{INTRODUCTION}

The density dependence of nuclear symmetry energy is crucial to
understand the structure of exotic nuclei, dynamics of heavy-ion
collisions, and many important issues in nuclear astrophysics such
as neutron star cooling and supernova explosive
\cite{liba2008,VBaran2005,JM.Lattimer2004,Steiner2005}. The ratio
$\pi^-/\pi^+$ was first proposed by \emph{Li} as a sensitive
observable for the high density behavior of the symmetry energy
\cite{liba2002}. After that, a lot of studies on $\pi^-/\pi^+$
ratio and the symmetry energy were carried out
\cite{lyz05,yong06,liq2006,Ferini06,gao11,gao13,XiaoZG:2009,FengZQ:2010,xie2013}.
However, recently different transport models, using the
$\pi^-/\pi^+$ ratio to interpret the FOPI pion data \cite{FOPI},
give contradict conclusions for the stiffness of the symmetry
energy \cite{gwm14}. Such situation stimulates more detailed
studies on pion production, such as different considerations of
the medium effects on the pion production, including energy
conservation on pion production \cite{Ferini06,XuJun13,Ko10,
cozma14} and pion potential in medium \cite{JunH14}. The effects
of pion potential \cite{ME66,CG88,Toki89,JunH14,EOset,Buss} on
pion production are in fact rarely studied in heavy-ion collision
around the pion production threshold. It is thus necessary to
investigate how the pion potential affects pion production since
some experimental studies on pion production are doing at
facilities that offer fast radioactive beams, such as the National
Superconducting Cyclotron Laboratory (NSCL) and the Facility
forRare Isotope Beams (FRIB) in the USA, or the Radioactive
Isotope Beam Facility (RIBF) in Japan.

\section{The IBUU Model and the pion potential}
\begin{figure}[t]
\centering
\includegraphics[width=0.9\textwidth]{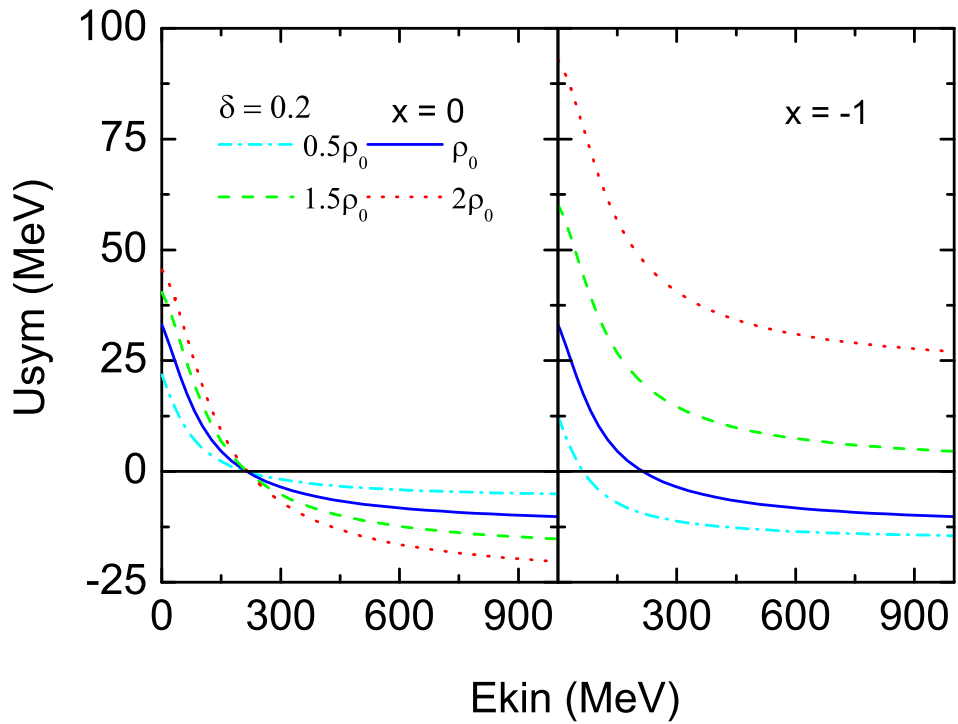}
\caption{(Color online) Density- and momentum-dependent symmetry
potential as a function of nucleonic kinetic energy with parameter
x = 0, -1. } \label{usym}
\end{figure}
In this study, we adopt the semi-classical transport model IBUU,
in which an isospin- and momentum-dependent mean-field single
nucleon potential is used \cite{Das03}, i.e.,
\begin{eqnarray}
U(\rho,\delta,\textbf{p},\tau)&=&A_u(x)\frac{\rho_{\tau'}}{\rho_0}+A_l(x)\frac{\rho_{\tau}}{\rho_0}\nonumber\\
& &+B(\frac{\rho}{\rho_0})^{\sigma}(1-x\delta^2)-8x\tau\frac{B}{\sigma+1}\frac{\rho^{\sigma-1}}{\rho_0^\sigma}\delta\rho_{\tau'}\nonumber\\
& &+\frac{2C_{\tau,\tau}}{\rho_0}\int
d^3\,\textbf{p}'\frac{f_\tau(\textbf{r},\textbf{p}')}{1+(\textbf{p}-\textbf{p}')^2/\Lambda^2}\nonumber\\
& &+\frac{2C_{\tau,\tau'}}{\rho_0}\int
d^3\,\textbf{p}'\frac{f_{\tau'}(\textbf{r},\textbf{p}')}{1+(\textbf{p}-\textbf{p}')^2/\Lambda^2},
\label{buupotential}
\end{eqnarray}
where $\tau=1/2(-1/2)$ for neutrons(protons),
$\delta=(\rho_n-\rho_p)/(\rho_n+\rho_p)$ is the isospin asymmetry,
and $\rho_n$, $\rho_p$ denote neutron and proton densities,
respectively. The parameter values $A_u(x)$, $A_l(x)$, $B$,
$C_{\tau,\tau}$, $C_{\tau,\tau'}$ $\sigma$, and $\Lambda$ can be
found in Ref. \cite{liba04}. $f_{\tau}(\textbf{r},\textbf{p})$ is
the phase-space distribution function at coordinate \textbf{r} and
momentum \textbf{p}. Different $x$ parameters can be used to mimic
different forms of the symmetry energy predicted by various
many-body theories without changing any property of the symmetric
nuclear matter and the symmetry energy at normal density. In fact,
what is particularly interesting and important for nuclear
reactions induced by neutron-rich nuclei is the isovector
(symmetry) potential. The present used symmetry potential $Usym =
(U_{n}-U_{p})/2\delta$, as shown in Figure~\ref{usym}, fits the
Lane potential data quite well \cite{chen07}.

\begin{figure*}[htb]
\centering\emph{}
\includegraphics[width=0.99\textwidth]{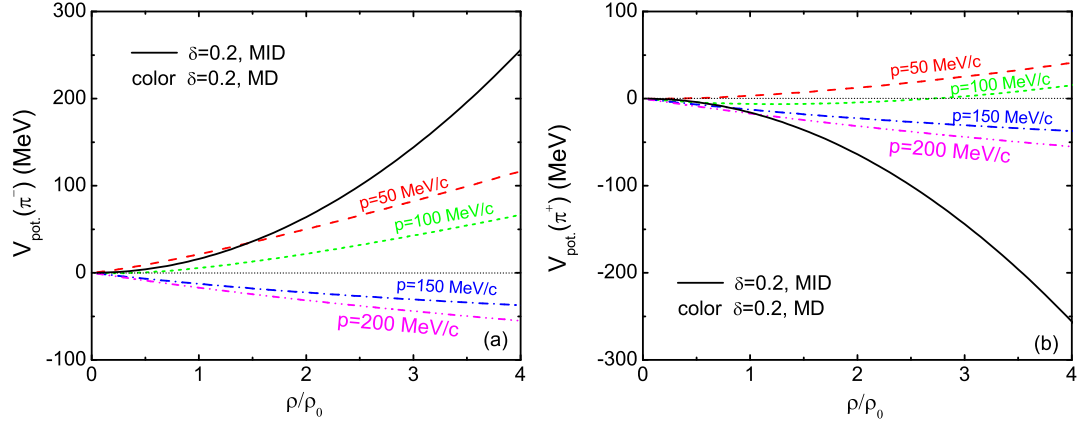}
\caption{(Color online) Density- and momentum-dependent pion
potential. The colored lines labelled by different momenta (MD)
are Buss's
potential \cite{Buss}. The black solid line denotes momentum-%
independent (MID) potential given by Jun Hong et al.
\cite{JunH14}. The left window shows negatively charged pion
potential while the right window shows positively charged pion
potential.} \label{Vpot}
\end{figure*}
\begin{figure*}[htb]
\centering
\includegraphics[height=6.8cm,width=0.40\textwidth]{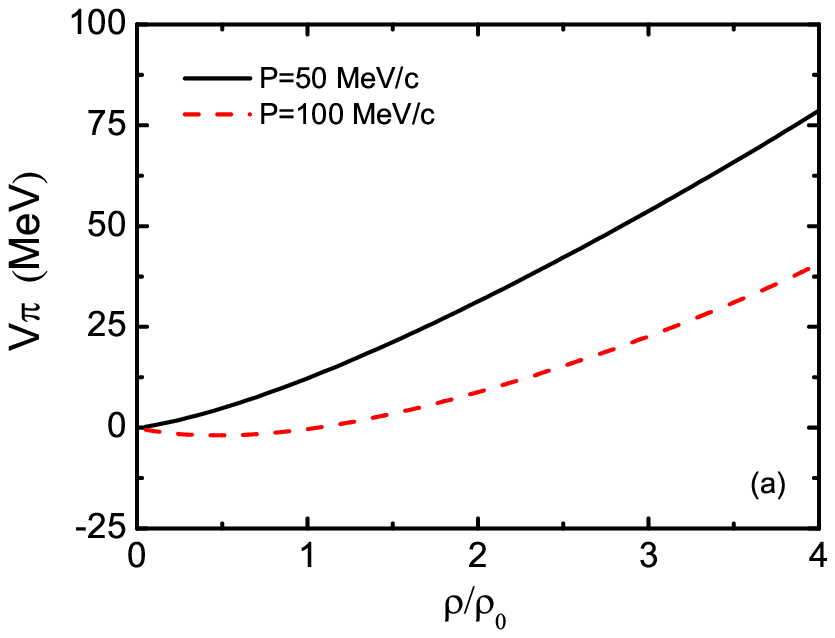}
\includegraphics[height=6.8cm,width=0.50\textwidth]{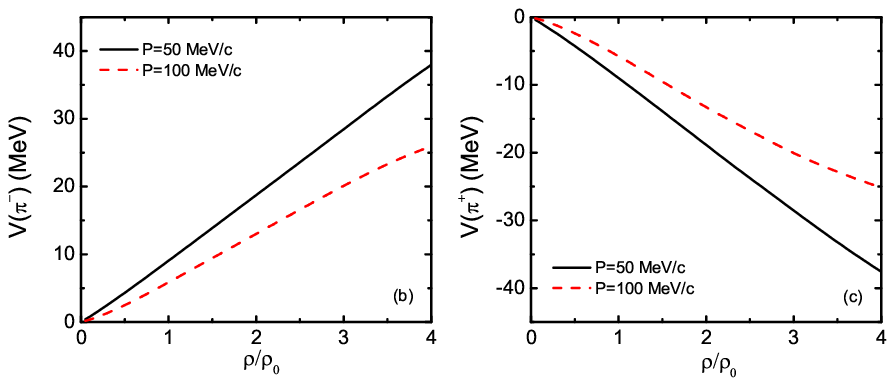}
\caption{(Color online) Isoscalar (Left) and isovector (Right)
potentials of pion (only $\delta= 0.2$ case is shown) used in the
IBUU model. } \label{svpion}
\end{figure*}
Pion-nucleus optical potential has been used to explain the
existence of pionic atoms. A pion potential had been constructed
by Toki et al., which can successfully describe the deeply bound
states of pionic atoms \cite{Toki89}. And Jun Hong et al.
constructed an isospin dependent pion potential \cite{JunH14}
which writing as $U_{\pi^{\pm}}=
\pm8S_{int0}\rho_T\frac{\rho^{\gamma-1}}{\rho^{\gamma}_0}$, where
$\rho_T$ is the isospin density ($\rho_{p}-\rho_{n}$)/2. This
potential agrees well with Toki's pion optical potential
\cite{Toki89}, but it only represents the so called s-wave
contribution to the $\pi$-nucleus potential. Shown in
figure~\ref{Vpot}, the black solid line, which is labelled by MID
is the pion potential constructed by Jun Hong et al. Here we
choose the parameter $\gamma$= 2 and $S_{int0}$= 20 MeV
\cite{JunH14}. It is seen that it is strongly density dependent,
repulsive for the $\pi^{-}$ and attractive for the $\pi^{+}$.

Figure~\ref{Vpot} also shows the other pion potential labelled by
colored lines MD, including so-called s-wave and p-wave
contributions, which divides into three parts according to
different regions of pionic momentum. In the low energy regime of
$p_{\pi} <$ 80 MeV, the results by Oset et al. are used
\cite{EOset,CG88,ME66}. When $p_{\pi} >$ 140 MeV, we use the form
of pion potential based on the $\Delta-$hole model \cite{Buss}. O.
Buss also constructed an appropriate potential when 80 MeV $\leq$
$p_{\pi} \leq$ 140 MeV by an interpolating spline to gain a
continuous derivative and therefore a continuous velocity of the
pion \cite{Buss}. From figure~\ref{Vpot}, we can clearly see that
this potential is density- and momentum-dependent, is repulsive at
low pionic momenta but becomes attractive at higher pionic
momenta. It is noted from figure~\ref{Vpot} that the ``MID'' pion
potential is just the isovector potential whereas the ``MD'' pion
potential includes both isoscalar and isovector potentials. The
latter is clearly shown in figure~\ref{svpion}. Because the
isoscalar potential is overall positive and the isovector
potential is positive for $\pi^{-}$ while it is negative for
$\pi^{+}$, and also the ``MD'' isovector potential is overall not
stronger than that of ``MID'' pion (isovector) potential, one sees
the two pion potentials are quite different. In this article,
based on the IBUU model, we use the second form of the pion
potential to study effects of pion potential and nuclear symmetry
energy in heavy-ion collisions at beam energies around the pion
production threshold.

\section{Results and discussions}

\begin{figure*}[htb]
\centering
\includegraphics[height=6.8cm,width=0.45\textwidth]{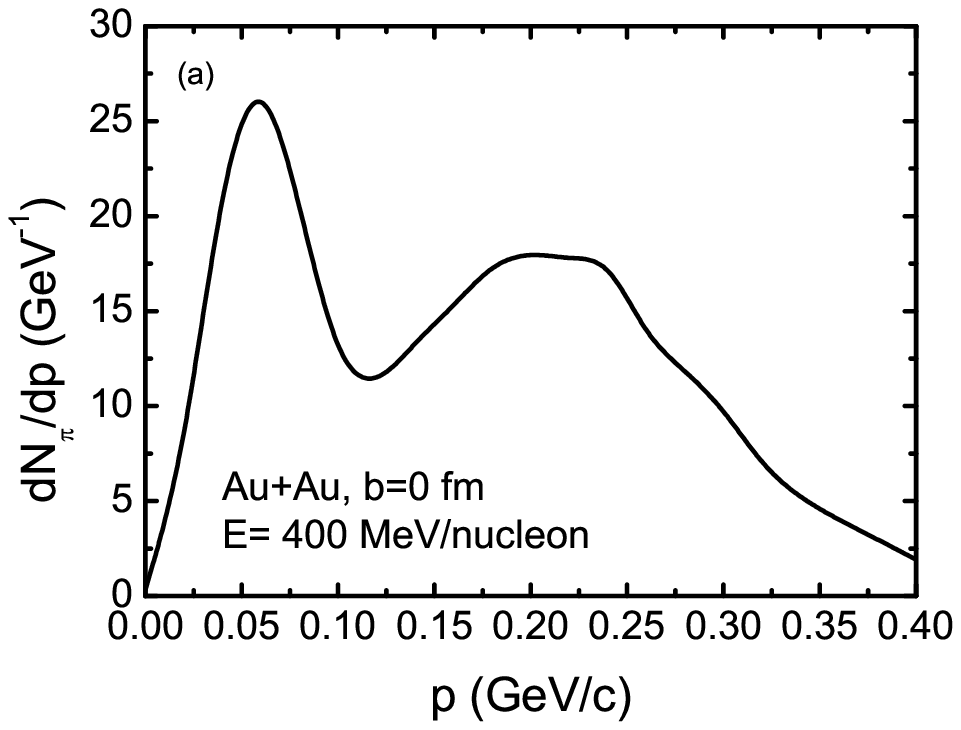}
\includegraphics[height=6.8cm,width=0.45\textwidth]{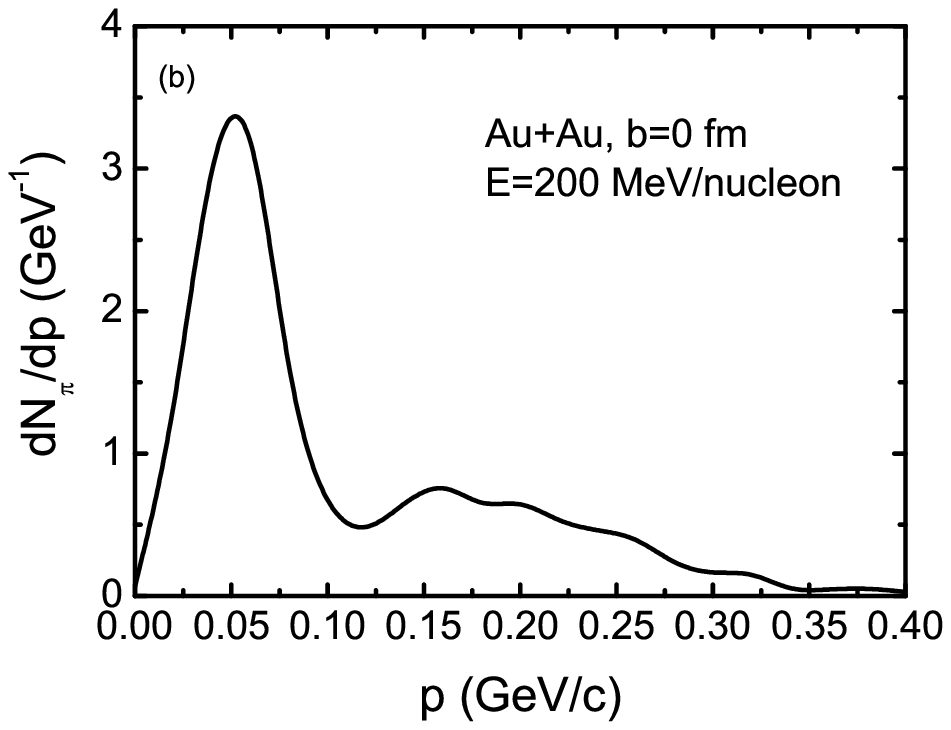}
\caption{Left: Momentum distribution of $\pi$ meson in the
compression stage in the central $^{197}Au+^{197}Au$ at
$E_{beam}$= 400 MeV/nucleon. Right: Same as the Left, but for
$E_{beam}$= 200 MeV/nucleon.} \label{dpion}
\end{figure*}
Since we use the density- and momentum-dependent pion potential,
it is necessary to show the momentum distribution of $\pi$ meson
in heavy-ion collisions. Figure~\ref{dpion} shows the momentum
distribution of $\pi$ meson in the central $^{197}Au+^{197}Au$ at
$E_{beam}$= 400 and 200 MeV/nucleon, respectively. Because at the
final stage of a reaction (the local baryon density of a $\pi$
meson is zero) $\pi$ suffers no pion potential, we in
figure~\ref{dpion} just show the momentum distribution of $\pi$
mesons in the compression stage. From the left panel of
figure~\ref{dpion}, it is shown that for the $^{197}Au+^{197}Au$
at $E_{beam}$= 400 MeV/nucleon, $\pi$ mesons are mainly located
around p= 58 MeV/c and p= 215 MeV/c. While for $E_{beam}$= 200
MeV/nucleon case, $\pi$ mesons are mainly located around p= 51
MeV/c. In the compressed stage, pion in the nuclear matter can be
absorbed by surrounding matter at certain momentum of pion meson,
thus one can see a sunken shape of pion distribution in momentum
space. This sunken shape is mainly decided by the kinematics of
pion as well as its absorption mechanism in nuclear matter while
it is less affected by the mean-field potential of pion. Different
momentum distributions of the $\pi$ mesons at different incident
beam energies should cause different effects of pion potential on
the charged pion ratio.

\begin{figure*}[htb]
\centering
\includegraphics[height=8.8cm,width=0.9\textwidth]{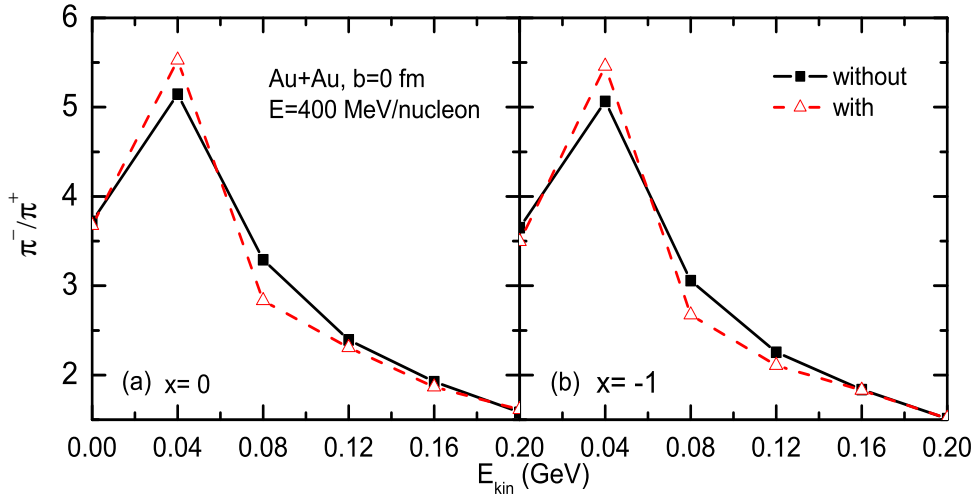}
\caption{(Color online) Left: Effect of pion potential on charged
pion ratio $\pi^{-}/\pi^{+}$ in the central $^{197}Au+^{197}Au$ at
$E_{beam}$= 400 MeV/nucleon with the softer symmetry energy (x=
0). Right: Same as the Left, but for the stiffer symmetry energy
(x= -1).} \label{effect400}
\end{figure*}
Figure~\ref{effect400} shows the effect of pion potential on
charged pion ratio $\pi^{-}/\pi^{+}$. It is seen that whether for
the softer symmetry energy or the stiffer symmetry energy, around
the Coulomb peak \cite{yong06}, the value of $\pi^{-}/\pi^{+}$
ratio with pion potential is higher than that without pion
potential. But for energetic pion mesons, a reversed case occurs.
The reason is that for $\pi^{-}$ with low energy, $\pi^{-}$ meson
suffers a repulsive pion potential (as shown in
figure~\ref{Vpot}). $\pi^{-}$ mesons are less absorbed thus more
$\pi^{-}$ production. Due to the Coulomb action, $\pi^{+}$ mesons
are less affected by the pion potential. Therefore we see a high
value of $\pi^{-}/\pi^{+}$ ratio around the Coulomb peak. For the
energetic $\pi^{-}$ mesons, they suffers an attractive pion
potential thus more $\pi^{-}$ mesons are absorbed by surrounding
nuclear matter. Then a low value of $\pi^{-}/\pi^{+}$ ratio is
seen.

\begin{figure*}[htb]
\centering
\includegraphics[height=8.8cm,width=0.9\textwidth]{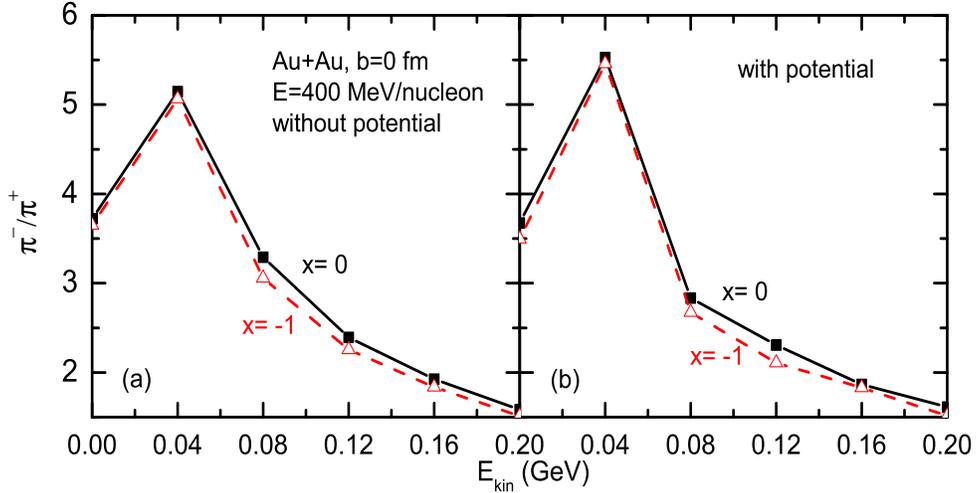}
\caption{(Color online) Effect of the symmetry energy on the
$\pi^{-}/\pi^{+}$ ratio in the central $^{197}Au+^{197}Au$ at
$E_{beam}$= 400 MeV/nucleon without (Left) and with (Right) the
pion potential.} \label{sympion400}
\end{figure*}
\begin{figure}[htb]
\centering
\includegraphics[height=8.8cm,width=0.8\textwidth]{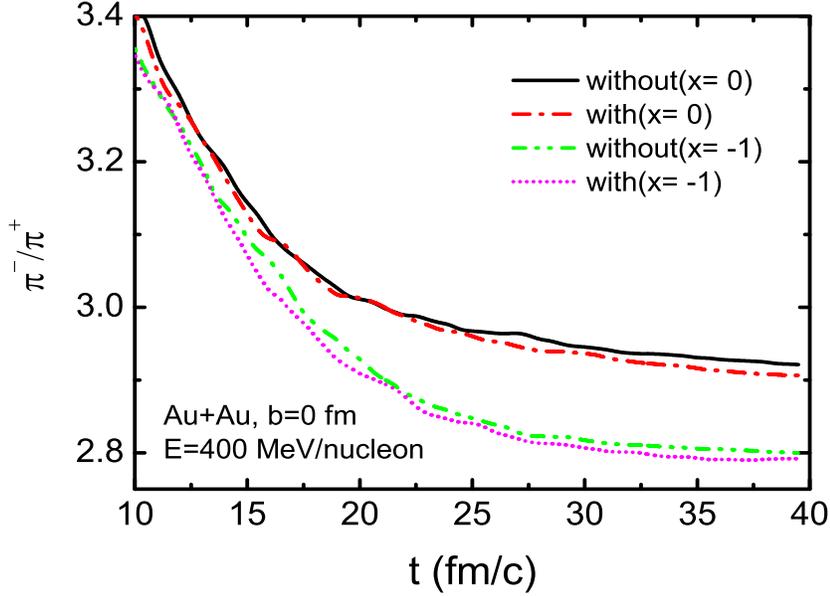}
\caption{(Color online) Time evolution of the $\pi^{-}/\pi^{+}$
ratio with and without pion potential for the softer (x= 0) and
the stiffer (x= -1) symmetry energies.} \label{time400}
\end{figure}
To answer whether the effect of the symmetry energy on the
$\pi^{-}/\pi^{+}$ ratio is affected by the pion potential, we plot
figure~\ref{sympion400}. It is seen that the effect of the
symmetry energy on the $\pi^{-}/\pi^{+}$ ratio is not affected
much by the pion potential at incident beam energy of 400
MeV/nucleon. This is more clearly shown in the
figure~\ref{time400}, time evolution of the $\pi^{-}/\pi^{+}$
ratio with and without pion potential for softer (x= 0) and
stiffer (x= -1) symmetry energies. From the figure~\ref{time400},
we see that total $\pi^{-}/\pi^{+}$ ratio is clearly affected by
the symmetry energy. The pion potential has less effect on the
$\pi^{-}/\pi^{+}$ ratio. As pointed out in the figure~\ref{dpion},
at incident beam energy of 400 MeV/nucleon, $\pi$ mesons are
mainly located around p= 58 MeV/c and p= 215 MeV/c. Therefore some
pion mesons suffer repulsive pion potential, but others suffer
attractive pion potential (shown in the figure~\ref{Vpot}). This
counterbalance makes pion potential has less effect on the
$\pi^{-}/\pi^{+}$ ratio at  $E_{beam}$= 400 MeV/nucleon, thus the
effect of the symmetry energy on $\pi^{-}/\pi^{+}$ ratio is kept.

\begin{figure*}[htb]
\centering
\includegraphics[height=8.8cm,width=0.9\textwidth]{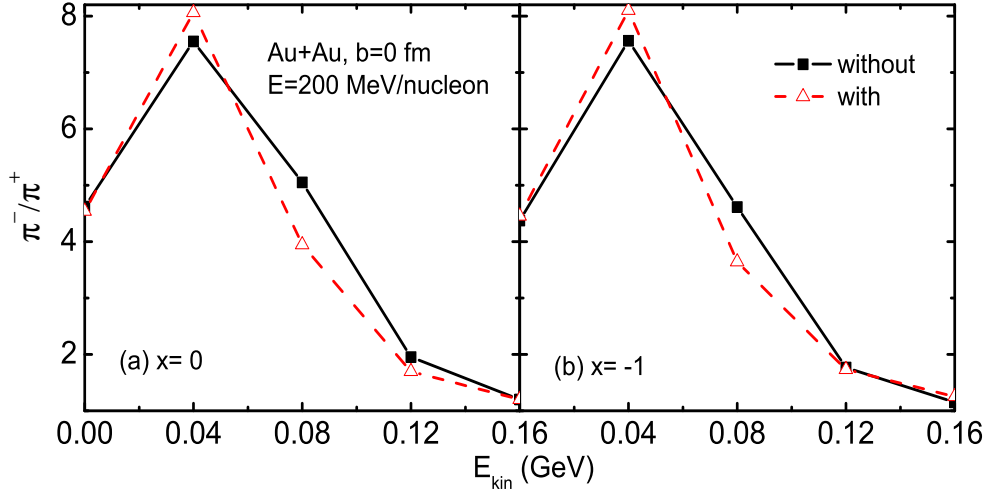}
\caption{(Color online) Same as Figure~\ref{effect400}, but for
$E_{beam}$= 200 MeV/nucleon.} \label{effect200}
\end{figure*}
\begin{figure*}[htb]
\centering
\includegraphics[height=8.8cm,width=0.9\textwidth]{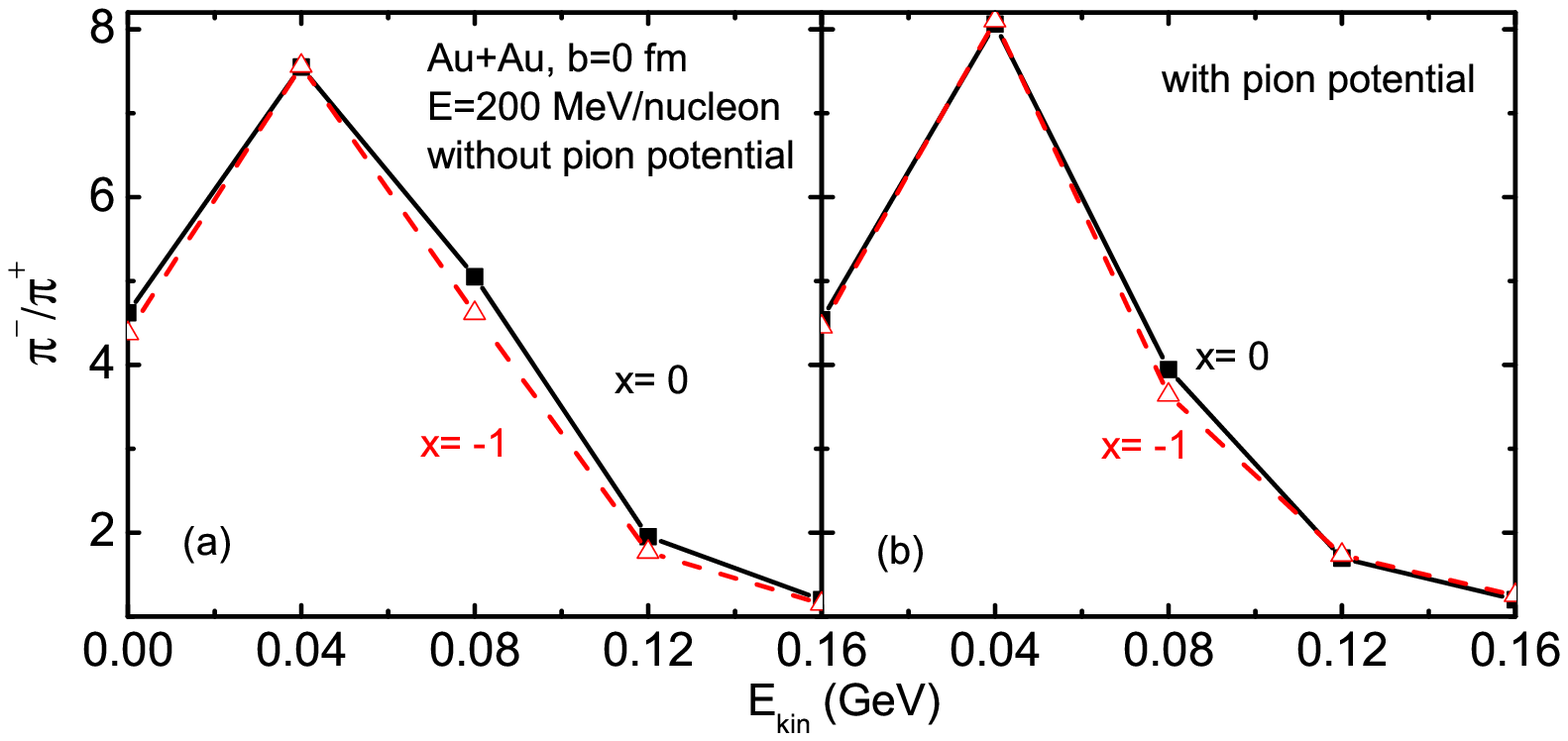}
\caption{(Color online) Same as Figure~\ref{sympion400}, but for
$E_{beam}$= 200 MeV/nucleon.} \label{sympion200}
\end{figure*}
\begin{figure}[htb]
\centering
\includegraphics[height=8.8cm,width=0.8\textwidth]{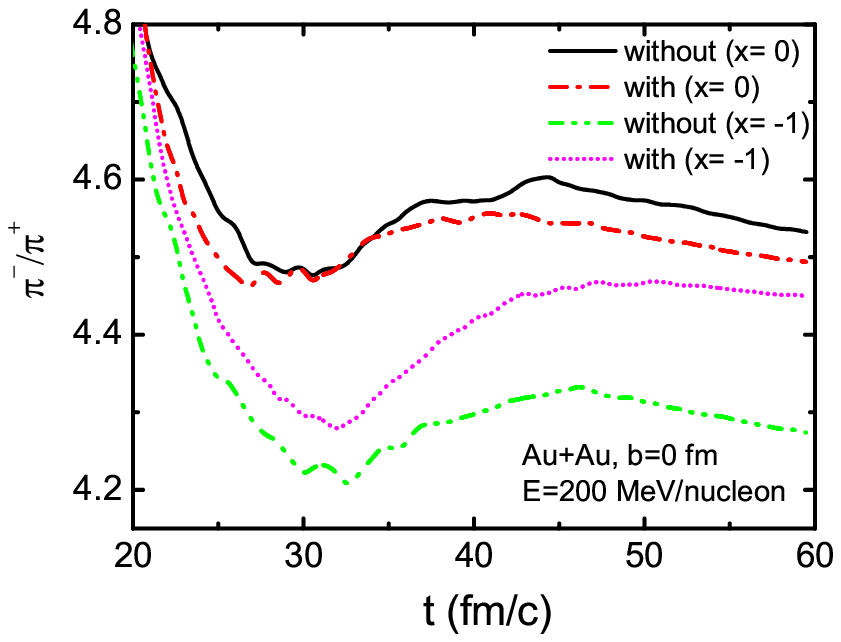}
\caption{(Color online) Same as Figure~\ref{time400}, but for
$E_{beam}$= 200 MeV/nucleon.} \label{time200}
\end{figure}
Since such related experiments are doing at NSCL/MSU ($E_{beam}$
$<$ 150 MeV/nucleon) and RIKEN/Japan ($E_{beam}$ = 200 $\sim$ 350
MeV/nucleon), it is necessary to show effects of pion potential
and nuclear symmetry energy on the $\pi^{-}/\pi^{+}$ ratio in
heavy-ion collisions at beam energies below the pion production
threshold. Shown in figure~\ref{effect200} is the effect of pion
potential on the $\pi^{-}/\pi^{+}$ ratio at the incident beam
energy of 200 MeV/nucleon with the softer (x= 0) and the stiffer
(x= -1) symmetry energies. Again, around the Coulomb peak and in
high pionic energy region the pion potential affects the
$\pi^{-}/\pi^{+}$ ratio much. And the effect of pion potential on
the $\pi^{-}/\pi^{+}$ ratio at $E_{beam}$= 200 MeV/nucleon seems
larger than that at $E_{beam}$= 400 MeV/nucleon. This is
understandable since at $E_{beam}$= 200 MeV/nucleon, pion mesons
are only located around p= 51 MeV/c where only a repulsive pion
potential dominates. To see how the large effect of pion potential
on the $\pi^{-}/\pi^{+}$ ratio at such beam energy affect the
probing of the symmetry energy by the $\pi^{-}/\pi^{+}$ ratio, we
plot figure~\ref{sympion200}, i.e., effect of the symmetry energy
on the $\pi^{-}/\pi^{+}$ ratio without and with pion potential. It
is seen that with pion potential, effect of the symmetry energy on
the $\pi^{-}/\pi^{+}$ ratio almost disappears. This is also
clearly seen in the figure~\ref{time200}, time evolution of the
$\pi^{-}/\pi^{+}$ ratio with and without pion potential for the
softer (x= 0) and the stiffer (x= -1) symmetry energies. As
demonstrated in Ref.~\cite{JunH14}, the large asymmetry
$(\rho_{n}-\rho_{p})/\rho$ of nuclear matter causes a large pion
potential. A larger pion potential for pion mesons with lower
energy causes $\pi^{-}$ mesons are less absorbed by surrounding
matter thus more $\pi^{-}$ mesons are produced finally. However,
on the other hand the large asymmetry of the nuclear matter also
causes large repulsion to neutrons by the symmetry potential, thus
less neutron-neutron collisions in matter make less $\pi^{-}$
mesons production. Therefore, the pion potential cancels out the
effect of symmetry energy on the $\pi^{-}/\pi^{+}$ ratio in
heavy-ion collisions at beam energies below the pion production
threshold.

\section{Conclusions}

In conclusions, we studied the effects of the pion potential and
the symmetry energy on the $\pi^-/\pi^+$ ratio in the central
$^{197}Au+^{197}Au$ collisions around the pion production
threshold. Because the symmetry potential in nuclear matter
generally repels neutrons out of the dense matter, which causes
small number of the neutron-neutron collisions. Small number of
neutron-neutron collisions causes small number of $\pi^-$ to
produce. However, the pion potential generally repels $\pi^-$ out
of the dense matter thus $\pi^-$'s are less absorbed. In heavy-ion
collisions, the pion potential causes large number of $\pi^-$'s
whereas the symmetry potential causes small number of $\pi^-$'s.
The effects of the symmetry energy on the $\pi^-/\pi^+$ ratio thus
cancelled out by the pion potential in some degree.

\section*{Acknowledgments}

The author G.C. Yong thanks Prof. Eulogio Oset and Dr. Jun Hong
for communications. The authors acknowledge the Texas Advanced
Computing Center (TACC) at The University of Texas at Austin for
providing HPC resources that have contributed to the research
results reported within this paper. This work is supported by the
National Natural Science Foundation of China (11375239, 11435014,
11175219). The 973 Program of China under No. 2013CB834405, the
Knowledge Innovation Project (KJCX2-EW-N01) of Chinese Academy of
Sciences.

\end{document}